\documentclass[superscriptaddress, showpacs, aps, twocolumn, pra, floatfix, 10pt]{revtex4-1}

\usepackage[utf8]{inputenc}

\usepackage[T1]{fontenc}
\usepackage{graphicx}
\usepackage{amsmath,amssymb,amsfonts,mathrsfs}
\usepackage[percent]{overpic}
\usepackage{subfigure}
\usepackage{braket}
\usepackage{bm}
\usepackage[breaklinks=true,colorlinks,citecolor=blue,linkcolor=blue,urlcolor=blue]{hyperref}
\usepackage{verbatim}

\begin{document}
\title{Unveiling operator growth in SYK quench dynamics}

\author{Matteo Carrega}
\thanks{All the authors contributed equally}
\affiliation{NEST, Istituto Nanoscienze-CNR and Scuola Normale Superiore, I-56127 Pisa, Italy}
\email[E-mail: ]{matteo.carrega@nano.cnr.it}

\author{Joonho Kim}
\thanks{All the authors contributed equally}
\affiliation{Institute for Advanced Study, Princeton, NJ 08540, USA}
\email[E-mail: ]{joonhokim@ias.edu}
\author{Dario Rosa}
\thanks{All the authors contributed equally}
\affiliation{School of Physics, Korea Institute for Advanced Study,
85 Hoegiro Dongdaemun-gu, Seoul 02455, Republic of Korea}
\email[E-mail: ]{dario85@kias.re.kr}

\begin{abstract}
We study non-equilibrium dynamics induced by a sudden quench of strongly correlated Hamiltonians with all-to-all interactions. By relying on a Sachdev-Ye-Kitaev
(SYK) based quench protocol, we show that the time evolution of simple spin-spin correlation functions is highly sensitive to the degree of locality of the corresponding operators, once an appropriate set of fundamental fields is identified. By tracking the time-evolution of specific spin-spin correlation functions and their decay, we argue that it is possible to distinguish between operator hopping and operator growth dynamics; the latter being a hallmark of quantum
chaos in many-body quantum systems. Such observation, in turn, could constitute a promising tool to probe the
emergence of chaotic behavior, rather accessible in state-of-the-art quench setups.
\end{abstract}

\maketitle


\section{Introduction}
\label{sec:introduction}

The study of strongly correlated quantum systems dates back to the early stages of quantum mechanics, and it still represents one of the most intriguing and challenging subjects of research~\cite{Engel:2007zzb, PhysRevE.91.022714, BRACHMANN1997391, PhysRevLett.93.142002, doi:10.1080/00018732.2010.514702, RevModPhys.83.863, PhysRevA.86.063627}. Recent technological advances both in condensed matter \cite{RevModPhys.78.217, PhysRevLett.112.166801, Kamata_2014} and atomic physics \cite{Kinoshita:2006, Cheneau:2012, Bloch2008Quantum, RevModPhys.80.885, Bloch2012Quantum, Langen:2015} allowed the investigation of many-body physics at the nanoscale, or single atom level, both in equilibrium and non-equilibrium settings.
In particular, state-of-the-art experiments with ultracold atoms and trapped ions \cite{RevModPhys.83.863, Blat:2012, Richerme:2014} offer the possibility to engineer closed quantum systems with very high precision and to perform quantum quench protocols \cite{RevModPhys.83.863, Calzona_2017, Calzona_2017a}; where non-equilibrium dynamics can be measured in real time after a sudden variation of some parameters of interacting Hamiltonians. Notable results have been already achieved in the context of relaxation dynamics and equilibration properties of quantum many-body systems \cite{Langen:2015, Eisert:2015}. Interestingly, optical lattice designs can be implemented to simulate quantum systems in various dimensions, under the presence of both local and $q$-body photon-mediated interactions \cite{Bohnet:2016, Gambetta:2020, Mottl_2012, Sch_tz_2014, M_ller_2012} between different lattice sites.

The above-mentioned progresses have attracted attention \cite{Eisert:2015} due to their potential in testing thermalization hypotheses and related conjectures in strongly correlated systems, which exhibit intriguing connections with quantum chaos and black-hole physics \cite{franz2018, Banerjee_2017, Lunkin_2018, Danshita_2017, PhysRevX.7.031006}. Popular observables in this context are out-of-time-ordered correlators (OTOCs) \cite{Larkin1969QuasiclassicalMI, Maldacena_2016, Garttner:2017} which have been recently used to quantify the chaotic nature of a given quantum system \footnote{We notice that currently there are, at least,  two notions of quantum chaos, which are believed to apply at different time (or energy) scales. The so-called early-time quantum chaos, as measured by the OTOCs, applies at time scales shorter than the scrambling (or Ehrenfest) time. The late-time quantum chaos, which applies at much longer time scales (of order of the Thouless time), is instead based on the statistics of the energy levels of a given quantum Hamiltonian, according to the Bohighas-Giannoni-Schmit conjecture \cite{Bohigas:1983er}. The connection between the two notions is still not clear and it constitutes an active area of research \cite{Cotler:2017jue, Xu_2020}. In this paper, we will refer to the early-time quantum chaos only.}. In fact, due to the exponential growth of the OTOCs in time when dealing with chaotic dynamics, they have been linked to the Lyapunov exponents of classical chaos, thereby being proposed as a promising measure of quantum chaos \cite{Maldacena_2016, Garttner:2017}. They quantify scrambling, or fast spreading of an initial local perturbation across the system. The spread of information in chaotic systems is intimately related to the notion of operator growth \cite{Sekino_2008, Roberts_2015, Hosur:2015ylk,Roberts:2018mnp, Qi_2019}, \textit{i.e.} to the idea that, during time evolution, local operators develop into rather complicated operators with increased spatial support and non-locality, resembling the so-called butterfly effect.

Although a few pioneering cold atom experiments have reported the possibility to measure the OTOCs in specific systems \cite{Garttner:2017}, an implementation of the OTOCs remains  very hard since it demands the ability of measuring different operators involving time-reversal of many-body dynamics. Several alternative diagnostics of quantum chaos \cite{qi2019measuring, Vermersch_2019, Joshi_2020, sundar2020proposal} have been thus proposed to circumvent this difficulty. Among them, Ref.~\cite{qi2019measuring} proposed that quantum chaos can be inferred by inspecting the temporal evolution of the full probability distribution of randomly prepared initial states after a quantum quench.

Inspired by this approach, in this work we focus on a paradigmatic example of dynamical
system equipped with all-to-all interactions, \textit{i.e.} the Sachdev-Ye-Kitaev (SYK) model \cite{PhysRevLett.70.3339, Kitaev15, Maldacena:2016hyu}.
It describes a strongly correlated quantum system of Majorana or Dirac fermions with random, all-to-all, $q$-body interactions, with $q$ being an integer larger than or equal to $2$. In particular, when $q > 2$, both the energy-level statistics and time-evolution of the OTOCs indicate that non-local interactions and random disorders make the model highly chaotic \cite{Cotler:2016fpe, Garcia-Garcia:2016mno, Kitaev15, Maldacena:2016hyu}.

Early-time quantum chaos in the SYK model was studied, from a slightly different perspective than the study of the OTOCs, also in \cite{Roberts:2018mnp}. The authors thereof studied the operator growth dynamics of the SYK model, with a particular emphasis on how simple operators, \textit{i.e.} consisting of products between few Majorana fermions, evolve into more complicated operators that involve products between a stack of Majorana fermions.

In this paper, we aim to study how the operator growth dynamics, linked to the onset of quantum chaos,  can be detected by quantum quench protocols \cite{Rossini:2019nfu, rosa2019ultra}.
Under the specific protocol where the quench Hamiltonian takes the form of SYK models, with different values of $q$, we investigate the dynamics of different spin correlation functions in 1D lattice spin systems. We argue that these correlators, which are quite accessible in state-of-the-art quench experiments by exploiting local imaging of quantum gas microscopes \cite{Kan_sz_Nagy_2018}, contain useful information on the operator dynamics.

In particular, we demonstrate that the decay rate of these correlation functions can be traced back to the operator growth dynamics under the SYK${}_q$ quench Hamiltonian, depending on the specific value of $q$.

The paper is organized as follows. In section \ref{sec:models}, we introduce the quench protocol and the models under study.
In section \ref{sec:magnetic_susceptibility}, we examine the quenched evolution of connected spin-spin correlation functions.
Specifically, we observe that a seemingly innocent modification, {\it i.e.} of the orientation of a static external magnetic field coupled to the spins in an initial local Hamiltonian, strongly affects the quench dynamics. In section \ref{sec:k-locality}, we explain the above observation in terms of the dynamical evolution of the ``fundamental'' operators in the theory.
We introduce the notion of \textit{operator hopping} --- to be contrasted to the operator growth --- which controls the dynamics under the integrable SYK Hamiltonians at $q = 2$. In section \ref{sec:operator_growth}, instead, we consider the case of the chaotic SYK${}_4$ models. We find that the dynamics is governed by operator growth instead of operator hopping, as expected for chaotic systems, demonstrating how temporal evolution of spin-spin correlators reflects the nature of operator dynamics.
Section \ref{sec:conclusions} summarizes our main results and possible perspectives.

\section{Model and quench protocol}
\label{sec:models}

The main focus of this work  is the study of non-equilibrium properties of strongly correlated quantum many-body systems. To this end, we consider a quench protocol, where the interactions between single entities of a quantum system
are suddenly switched on at time $t=0$ as
\begin{equation}
\label{eq:quench}
{\cal \hat{H}}(t)= {\cal \hat{H}}_0 + \theta(t)[{\cal \hat{H}}^{(q)}_1 -{\cal \hat{H}}_0]~,
\end{equation}
where $\theta(t)$ is a step function and the dynamics for $t > 0$ is entirely governed by ${\cal \hat{H}}^{(q)}_1$ in a scale invariant fashion. Such quantum quench protocols can be engineered, and have been realized, for example in cold-atoms setups \cite{RevModPhys.83.863}.
The system is initially prepared (for $t<0$) in the ground state of a generic  free and local ${\cal \hat{H}}_0$, {\it i.e.} a quantum system on a spin lattice of length $L$, and then it evolves under the action of a strongly interacting Hamiltonian with all-to-all couplings.
For the latter, we consider the SYK${}_q$ Hamiltonian that represents a paradigmatic example of all-to-all interacting systems \cite{PhysRevLett.70.3339, Kitaev15, Maldacena:2016hyu, franz2018}. These models recently gained a lot of attention due to  intriguing connections with black-hole physics and since they can show  quantum chaotic behavior \cite{franz2018}.
This family of Hamiltonians, classified in terms of an integer parameter $q$, can be written in terms of $2L$ Majorana fermions, interacting via $q$-body and all-to-all coupling terms as  \cite{Maldacena:2016hyu, PhysRevLett.70.3339,Kitaev15}
\begin{align}
\label{eq:SYK_hamiltonians_def}
& \mathcal{\hat H}_{1}^{(q)} = (\mathrm{i})^{q / 2} \, \sum_{i_1 <  \dots < i_q} J_{i_1  \dots i_q} \hat \gamma^{i_1} \dots \hat \gamma^{i_q} \ .,
\end{align}
where the Majorana fermion operators, $\hat \gamma^i$, satisfy the following Clifford algebra relations:
\begin{equation}
\label{eq:majorana_algebra}
\{ \hat \gamma^i , \, \hat \gamma^j \} = \delta^{i j} \ ,
\end{equation}
and where the coupling constants $J_{i_1  \dots i_q}$  are extracted from a Gaussian distribution, with null mean value and variance
\begin{align}
    \label{eq:Jcouplings_variances}
\overline{J_{i_1\cdots i_q}^2}  &= \frac{J^2(q-1)!}{(2L)^{q-1}}.
\end{align}
Throughout the paper, we set $\hbar=1$ and the overline will denote the Gaussian average over all the SYK coupling constants $J_{i_1\cdots i_q}$.
Unless otherwise specified, we also set $J^2 = 1$.
While at $q=2$ the SYK model has been known to show integrable behavior \cite{Gross_2017}, the situation gets completely different for higher values of $q > 2$: although sharing all-to-all correlations, in this case, the model turns out to be chaotic based on the study of the
 OTOCs \cite{Kitaev15, Maldacena:2016hyu} as well as
 the energy-level statistics \cite{Cotler:2016fpe, Garcia-Garcia:2016mno}.
 It is thus interesting to take both the $q=2$ and $q>2$ models as the quench Hamiltonian, investigating how they differently affect the relaxation dynamics, looking for features related to the presence of quantum chaos.

For sake of simplicity, we consider the free and local ${\cal \hat{H}}_0$ as an ensemble of non-interacting spin-$\frac{1}{2}$ variables on a lattice of length $L$ immersed on a transverse magnetic field oriented along the $a=x,y,z$ direction
\begin{equation}
\label{eq:H_0_def}
\mathcal{\hat H}_0^{a} \equiv \sum_{i = 1}^L \hat h_i^{a} \equiv \omega \sum_{i = 1}^{L} \, \hat \sigma_i^{a} \ ,
\end{equation}
where $\hat \sigma_i^{a}$ denotes the $a$-th Pauli matrix at site $i$ and $ \omega$ is the magnetic field strength, hereafter assumed to be the same for all directions (and equal to $1$) without loss of generality.
The spin Hamiltonian defined in Eq.~\eqref{eq:H_0_def} can be mapped into a system of $2L$ Majorana fermions via the following Jordan-Wigner (JW) map \cite{sachdev_2011},
\begin{align}
\label{eq:JWmaps}
& \hat \gamma^{2j-1}  =  \frac{1}{\sqrt{2}} \left(\prod_{i=1}^{j-1} \hat \sigma^z_i \right) \hat \sigma^x_j \ , \nonumber \\
& \hat \gamma^{2j}   =  \frac{1}{\sqrt{2}} \left(\prod_{i=1}^{j-1} \hat \sigma^z_i \right) \hat \sigma^y_j \ ,
\end{align}
which defines the duality between $L$ spins and $2L$ Majorana fermions.
Obviously, also the SYK$_q$ Hamiltonians can be mapped to the spin chain variables via the JW map \eqref{eq:JWmaps}, but this would result in expressions that are highly non-local and not easy to put in a compact form.

The lattice spin Hamiltonians \eqref{eq:H_0_def} can be realized in several settings, such as cold atoms, trapped ions or solid state devices.
For example, one can engineer a system of neutral atoms with hyperfine interactions and coupling them to Rydberg states \cite{Labuhn_2016,Jau_2015,PhysRevX.7.041063,Bernien_2017}.
A transverse field can be then  introduced by applying a resonant microwave or Raman coupling between two hyperfine states, and several type of correlations, ranging from local to all-to-all interactions, have been proposed and recently realized, exploiting photon-mediated correlations in optical cavities as well \cite{Bohnet:2016, Gambetta:2020}. In passing, we mention that some recent proposals have suggested the possibility to realize SYK-like Hamiltonians in all these settings \cite{franz2018, Danshita_2017, Chew_2017, Chen_2018}.

We are interested in tracing the time-evolution of the ground states, $\ket{0^a}$, after the action of the quench protocol of Eq.~\eqref{eq:quench}. In order to make fair comparisons of time-evolution dynamics across different spin models with various sizes and interaction types, the quench Hamiltonian can be properly normalized as ${\cal \hat{H}}\to \mathscr{\hat H}^{(q)}_1$ to have a unit bandwidth.
This can be done \cite{Rossini:2019nfu} by renormalizing all coupling constants by the energy bandwidth of the interaction Hamiltonian, $\Delta_{\mathcal{\hat H}^{(q)}_1}$, defined as the energy gap between the maximum/minimum eigenvalues, \textit{i.e.}
\begin{equation}
\label{eq:H1_renormalized}
\mathscr{\hat H}^{(q)}_1 \equiv
\frac{ \mathcal{\hat H}^{(q)}_1}{\Delta_{\mathcal{\hat H}^{(q)}_1}}
= \frac{ \mathcal{\hat H}^{(q)}_1}{E_{\mathcal{\hat H}^{(q)}_1}^{\rm{Max}} - E_{\mathcal{\hat H}^{(q)}_1}^{\rm{Min}}}.
\end{equation}

In the following, in line with the recent results of \cite{qi2019measuring}, we show that the SYK quench Hamiltonian can produce markedly distinct effects on the dynamics starting from different initial states and depending on the operator dynamics under investigation.
To this end we will focus on two possible choices, $a=x$ or $a=z$, for ${\cal \hat{H}}_0^{a}$ (the case $a=y$ being completely equivalent to the case $a=x$).

At first sight, the distinction between the two models may look very minor, since they can be related by a simple rotation of the external magnetic field. However, they have different dynamical features: for example, the global symmetries of the constant Hamiltonians \eqref{eq:H_0_def} are broken by the quench term in different ways.
Indeed, $\mathcal{\hat H}_0^{x}$ preserves the total spin along the $x$ axes and the quench term \textit{completely} breaks this symmetry
(the exactly same pattern happens for the $a = y$ Hamiltonian).
On the other hand, $\mathcal{\hat H}_0^{z}$ preserves the total spin along the $z$  axes and this symmetry is solely \textit{partially} broken by the quench term down to the parity symmetry.

\section{Dynamical spin-spin correlation functions}
\label{sec:magnetic_susceptibility}

Time-resolved detection of propagating correlations in an interacting quantum many-body system after a quantum quench have been recently reported \cite{Cheneau:2012, Langen:2015, Alba_2017, Bastianello_2018}.
In addition, the ability offered by quantum gas microscopy \cite{Haller_2015, Yamamoto_2016, Kan_sz_Nagy_2018}
of single-site imaging in an optical lattice \cite{Parsons_2016} allows the spatial resolution and sensitivity  to reveal the real-time evolution of a many-body system at the single-particle level.
Motivated by these progresses, we numerically investigate, by means of discrete time-evolution simulations, the dynamics of the evolved states after the quantum quench by studying the time evolution of the connected part of the two-point function or dynamical susceptibility,
\begin{equation}
\label{eq:magnetic_suscept}
\chi^{a}(t) \equiv \sum_{i < j} \left(\overline{\langle \hat \sigma_i^{a} \, \hat \sigma_j^{a} \rangle} - \overline{\langle \hat \sigma_i^{a} \rangle} \; \overline{\langle \hat \sigma_j^{a} \rangle} \right)\ ,
\end{equation}
where the expectation value $\langle \cdots \rangle$ is taken over the evolved ket, defined as
\begin{equation}
\label{eq:evolved_state}
\ket{\psi(t)^{a}} \equiv e^{- i \mathcal{\hat H } (t) \, t} \, \ket{0^{a}} \ .
\end{equation}
For the sake of simplicity, we start discussing the case of quench Hamiltonian of SYK$_2$ type, while larger values $q>2$ are considered in later sections.
We recall that, when $q = 2$, the SYK Hamiltonian is non chaotic, thus providing an interesting example of disordered, all-to-all, integrable dynamics.
\begin{figure}
    \centering
    \includegraphics[width=8cm
]{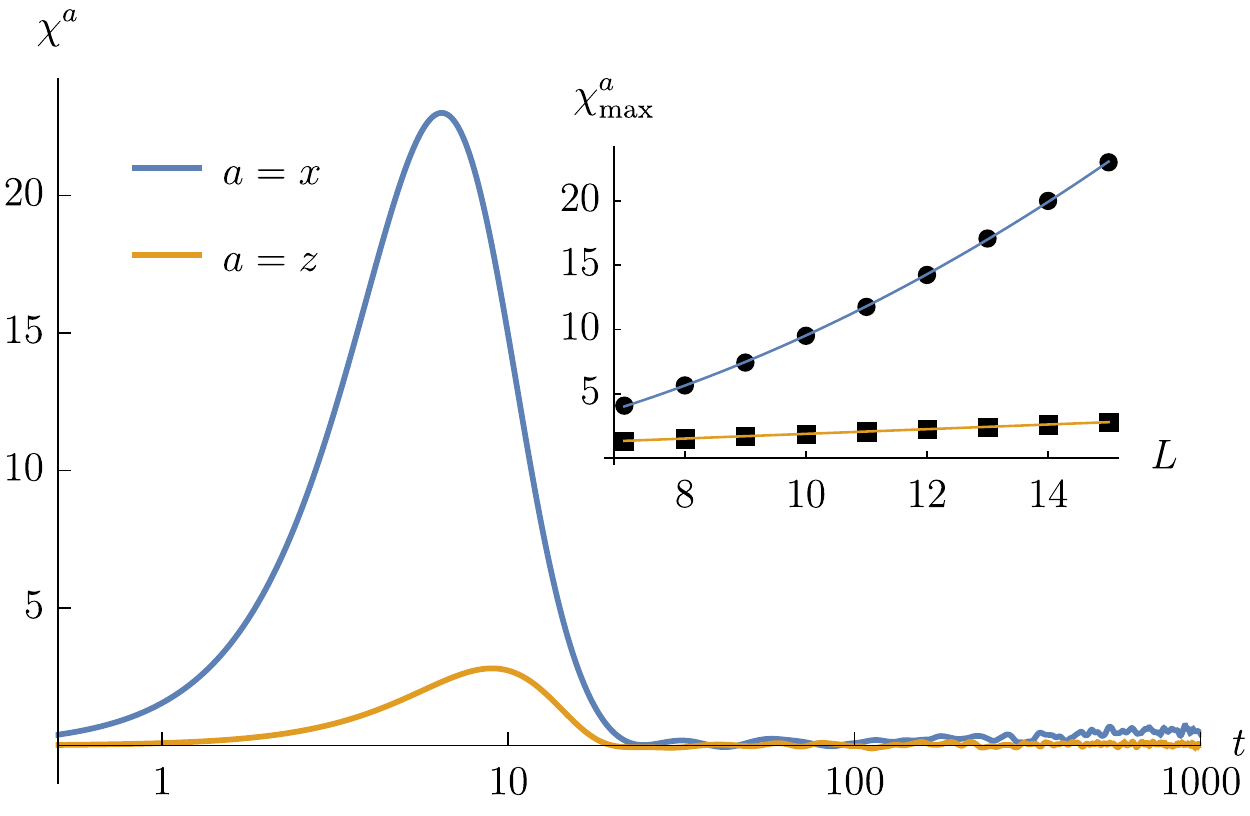}
    \caption{The function $\chi^a (t)$ for both the $a = x$ and the $a = z$ model computed for $L = 15$, averaged over $300$ ensemble realizations. Inset: the values of $\chi^a_\mathrm{max}$ as functions of $L$ (black circles, for $a = x$, and black squares, for $a = z$). We observe that these behaviours are very well reproduced by the functions $\chi^x_\mathrm{max} (L) = a L^2 + b$ (blue line) and $\chi^z_\mathrm{max} (L) = c L + d$ (orange line).
    The fitting parameters are $a = 0.11$, $b = -1.3$, $ c = 0.18$ and $d = 0.083$.}
    \label{fig:two_point}
\end{figure}
The time evolution of $\chi^a(t)$ for both $a = x, \, z $ is reported in Fig.~\ref{fig:two_point}.
In both cases, this quantity  shows an initial rise, up to a maximum value $\chi^{a}_{\mathrm{max}}$, followed by a decrease toward $0$.
Interestingly, the peak is drastically larger in the $a = x$ model, with the difference getting parametrically enlarged by increasing $L$.
More in details, we have observed that $\chi^x_\mathrm{max}$ grows quadratically with $L$, while $\chi^z_\mathrm{max}$ grows linearly with $L$ (See the inset in Fig~\ref{fig:two_point}).
Such difference is not obvious to be explained, given that the two pre-quench models as well as their correlators just differ  by a global rotation.

To better clarify these behaviors, it is instructive to inspect the single spin-spin correlation functions
\begin{equation}
\label{eq:magnetic_suscept_ij}
\chi_{i j}^{a}(t)\equiv \overline{\langle \hat \sigma_i^{a} \, \hat \sigma_j^{a} \rangle} - \overline{\langle \hat \sigma_i^{a} \rangle} \, \overline{ \langle \hat \sigma_j^{a} \rangle} \quad \mathrm{with} \; i < j \ ,
\end{equation}
appearing in the sums of Eq.~\eqref{eq:magnetic_suscept}, with $i,j$ indicating two lattice sites.
Across all $i, j$ combinations, the time-evolution of $\chi_{i j}^{a}(t)$ exhibits the same qualitative pattern as of $\chi^a (t)$, which is an initial rise followed by a decay.
This is because both $\overline{\langle \hat \sigma_i^{a} \, \hat \sigma_j^{a} \rangle}$ and $\overline{\langle \hat \sigma_i^{a} \rangle} \, \overline{\langle \hat \sigma_j^{a} \rangle}$ decay monotonically from one to zero but with different rates, such that $\chi^{a}_{ij}(t)$ and $\chi^{a}(t)$ develop a peaked structure.

To understand the differences in the peaks between the $a=x$ and $a=z$ models, we study the height of the peaks $P_{ij}^a \equiv \mathrm{max}_t( \chi^{a}_{ij}(t))$ as a function of $i$ and $j$. Naively, one would expect $P_{ij}^a$ to be homogeneous and independent of $i$ and $j$, because the initial Hamiltonian \eqref{eq:H_0_def} treats all the spins equally and does not involve spin-spin couplings, and the SYK term \eqref{eq:SYK_hamiltonians_def} is all-to-all.
Hence, one would expect that the models do not have any notion of neighboring sites, therefore  $P_{ij}^a$ would be independent of the lattice sites after averaging over the Gaussian random couplings.

\begin{figure}
    \centering
    \includegraphics[width=\linewidth]{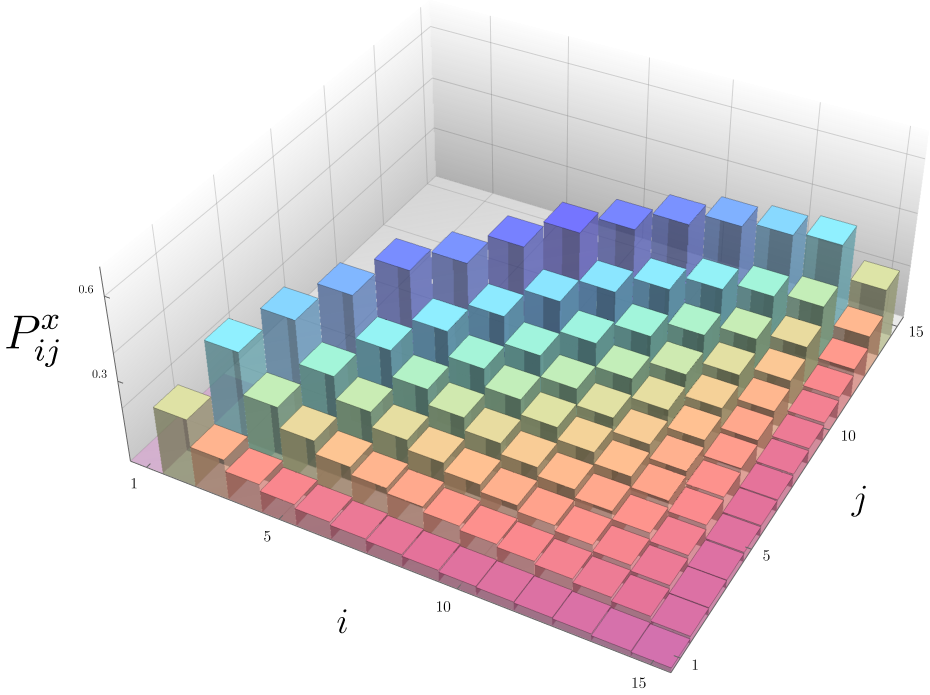}
    \includegraphics[width=\linewidth]{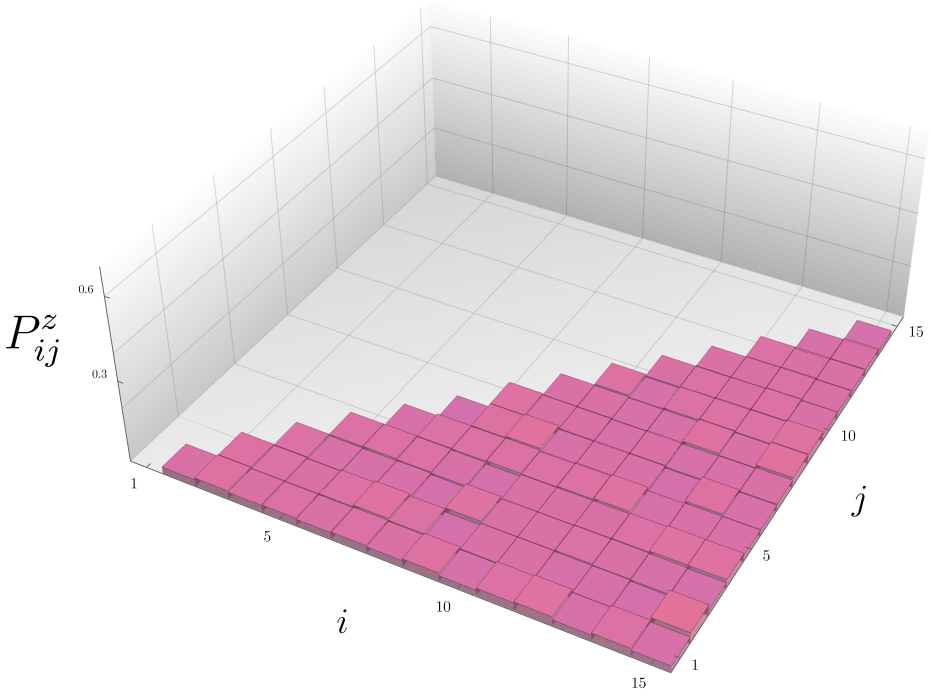}
    \caption{The peak heights of $P^x_{ij} $ and $P^z_{ij} $ as a function of the lattice sites $i,j$ at $L=15$ and averaged over $300$ ensemble realizations.}
    \label{fig:histogram}
\end{figure}
Numerical results on $P_{ij}^a$ are displayed in Fig.~\ref{fig:histogram}.
In agreement with the motivations just explained above,  $P_{ij}^z$ is independent of $i, \, j$.
On the other hand, $P_{ij}^x$ exhibits a very intriguing behavior; it is \textit{highly} dependent on $i$ and $j$.
In addition, the following patterns clearly emerge; the peaks are more pronounced for $i$ and $j$ being closer to each other, and they are further magnified when $i$ and $j$ approach to the center of the lattice.

Notice that these differences are not very easy to understand solely from the symmetry breaking perspective.
For example, it is not obvious why $P_{ij}^x$ is site-dependent.
Moreover, in App.~\ref{App:hard-core_bosons} we provide further evidences that the symmetry breaking pattern alone cannot explain the dynamical differences observed between ${\cal \hat{H}}_0^x$ and ${\cal \hat{H}}_0^z$, considering hard-core bosonic versions of the SYK models. Although the bosonic variants show the same symmetry breaking patterns as in their fermionic counterparts, they exhibit clearly distinct dynamical properties.

\section{Locality in the operator space and quantum dynamics}
\label{sec:k-locality}

Here we demonstrate how the difference between $P_{ij}^x$ and $P_{ij}^z$ presented above can be  explained by inspecting the temporal decay properties of the correlation functions $\overline{\langle \hat \sigma_i^{a} \, \hat \sigma_j^{a} \rangle}$ and $\overline{\langle \hat \sigma_i^{a} \rangle}$. Our approach is similar to the one discussed in \cite{Roberts:2018mnp} and based on the notion of operator size.

It is fairly simple to convert the Pauli matrices $\hat \sigma_i^a$ as well as the product $\hat \sigma_i^a \hat \sigma_j^a$ of any two Pauli pairs
into a product of Majorana fermions $\hat \gamma^i$ \cite{Kitaev:2001kla}, by making use of the JW maps defined in \eqref{eq:JWmaps}.
For this reason, we find more convenient to treat the Majorana fields $\hat \gamma^i$, rather than the Pauli operators $\hat \sigma_i^a$, as the \textit{fundamental} operators \cite{Roberts:2018mnp} to get a more transparent tracking of the time-evolution of the spin correlators.

Given the set of fundamental operators, one can introduce the notion of \textit{size} of operators: an operator $\hat{\mathcal{O}}$ is said to have size $k$ if it can be written as a product of $k$ fundamental operators. More generally, an operator is $k$-local if, once rewritten as a sum of operators with definite size, the maximum size of its constituents is $k$. We will denote the locality of an operator $\hat{\mathcal{O}}$ in the superscript, \textit{i.e.} $ \hat{ \mathcal{O}}^{(k)}$.
It has been emphasized in \cite{Roberts:2018mnp} that a proper identification of the set of fundamental operators, and the associated notion of operator size, allow a clear description of the dynamics of a strongly correlated quantum system.

Let us start considering the $a = z$ case, whose description is simpler.
 By making use of the JW map \eqref{eq:JWmaps}, the operators $\hat \sigma_i^{z}$ and $\hat \sigma_i^{z} \, \hat \sigma_j^{z}$ can be re-written in terms of the Majorana variables as follows:
\begin{align}
\label{eq:k-locality_z_system}
\begin{split}
& \hat \sigma_i^{z} = - 2 \mathrm{i} \, \hat \gamma^{2i - 1} \hat \gamma^{2i} \ ,  \\
& \hat \sigma_i^{z} \hat \sigma_j^{z} = - 4 \, \hat \gamma^{2i - 1} \hat \gamma^{2i} \hat \gamma^{2j - 1} \hat \gamma^{2j} \ ,
\end{split}
\end{align}
thus showing that they are, respectively, of size $2$ and $4$ in the space of the Majorana operators. It is noticeable that the sizes of the operators $\hat \sigma_i^{z}$ and $\hat \sigma_i^{z} \, \hat \sigma_j^{z}$ are \textit{independent} of the lattice sites $i$ and $j$.

The situation is more involved in the $a = x$ case. Here, the spin operators, $\sigma_i^x$, show varying sizes which \textit{depend} on the lattice site. More precisely, we find that $\hat \sigma_i^{x}$ is an operator of size $(2 i - 1)$, \textit{i.e.}
\begin{equation}
\label{eq:k-locality_x_system}
\hat \sigma_i^{x} =  \bigl( \sqrt{2} \bigr)^{2 i - 1} (- \mathrm{i})^{i - 1} \, \prod_{p = 1}^{2 i - 1} \hat \gamma^{p} \ .
\end{equation}
The application of the Clifford algebra \eqref{eq:majorana_algebra} shows similarly that the product operator $\hat \sigma_i^{x} \hat \sigma_j^{x}$ is of size $2|j-i|$. We summarize the size of the above spin operators as follows:
\begin{align}
\begin{tabular}{c|cccc}
     Operator
     & $\hat \sigma_i^{z}$
     & $\hat \sigma_i^{z} \hat \sigma_j^{z}$
     & $\hat \sigma_i^{x}$
     & $\hat \sigma_i^{x} \hat \sigma_j^{x}$ \\\hline
     Size
     & $2$
     & $4$
     & $2i-1$
     & $2|j-i|$
\end{tabular}
\label{eq:k-locality_x_system_summary}
\end{align}

The post-quench evolution of a generic spin operator $\mathcal{O}^{(k)}$ of size $k$ follows the Heisenberg equation of motion,
\begin{align}
    \label{eq:Heisenberg}
    \frac{d}{dt} \langle \mathcal{O}^{(k)}(t) \rangle = \mathrm{i} \langle [\mathscr{\hat H}^{(2)}_1, \mathcal{O}^{(k)}(t)] \rangle,
\end{align}
which involves the commutator between the SYK$_2$ Hamiltonian and the operator itself. An advantage of the Majorana representation comes from the fact that the above commutator can be computed in a very simple way that manifestly preserves the operator size, by using
\begin{equation}
\label{eq:operator_hopping_algebra}
\left[\mathcal{\hat H}_{1}^{(2)} , \hat \gamma^i \right] = - \mathrm{i} \sum_j J_{ij} \, \hat \gamma^j \ .
\end{equation}
Hence, the operator dynamics, in the space of Majorana operators, under the SYK$_2$ Hamiltonian is a kind of \textit{operator hopping}: an operator $\mathcal{O}^{(k)}$, under time evolution, moves along the space of the operators of the \textit{same} size $k$. The operator size does not grow in time. We underline that this characteristic is very distinct from the operator dynamics under SYK$_q$ Hamiltonians with $q > 2$, studied in \cite{Roberts:2018mnp} which will be discussed in the later section.

We are now at the position to explain the observed differences in the time evolution of the spin-spin correlation functions $\chi^a_{ij}(t)$, in terms of $\overline{\langle \hat \sigma_i^{a} \, \hat \sigma_j^{a} \rangle}$ and $\overline{\langle \hat \sigma_i^{a} \rangle}$, for the $a = x$ and $a = z$ cases. First of all, after averaging over the SYK coupling constants, the above correlators turn out to be \textit{return amplitudes}, \textit{i.e.} they compute the amplitude that at $t > 0$ the evolved operator takes exactly the same form of the initial operator.
This is a consequence of the Gaussian averaging, as we show explicitly in App.~\ref{App:gaussian_average_perturbative}.
\footnote{Contrary to the analysis of Ref.\cite{Roberts:2018mnp}, we are considering correlators which are \textit{averaged} over the Gaussian couplings $J_{i_1 \dots i_q}$; while the analysis of \cite{Roberts:2018mnp} is performed at fixed (but random) values of the couplings $J_{i_1 \dots i_q}$. When computed at fixed values of the couplings $J_{i_1 \dots i_q}$, the return amplitude for the fundamental operator $\hat{\gamma}^{i}$ is proportional to $\mathrm{Tr}(\hat \gamma^i(t) \, \hat \gamma^i)$.}

Secondly, when $t$ sufficiently increases, the averaged correlators should decay to $0$, since the initial operator is of measure $0$ in the entire space of size $k$ operators.

Given these facts, we argue that the vanishing of the spin-spin correlator happens \textit{faster} when the size $k$ of the corresponding operator is \textit{larger}, if $k \leq L$, and when the size $k$ is smaller, if $k>L$. Indeed, as $k$ becomes bigger, the dimension of the space of all size $k$ operators
\begin{align}
\label{eq:dim}
    \text{dim}_{k} =  \frac{(2L)!}{k!(2L-k)!}
\end{align}
increases monotonically until it reaches $k = L$, \textit{i.e.} exactly half of the maximum possible size for an operator, and then decreases up to $k = 2L$.
In short, $\hat{\mathcal O}^{(k)}(t)$ has a larger space to explore as $|k-L|$ decreases. This causes the corresponding return amplitude $\overline{\langle\hat{\mathcal O}^{(k)}(t)\rangle}$ to vanish faster, since the chance to go back to the original operator $\hat{\mathcal O}^{(k)}$ will get reduced in a shorter amount of time.

\begin{figure}[ht]
\centering
	\includegraphics[width=8cm]{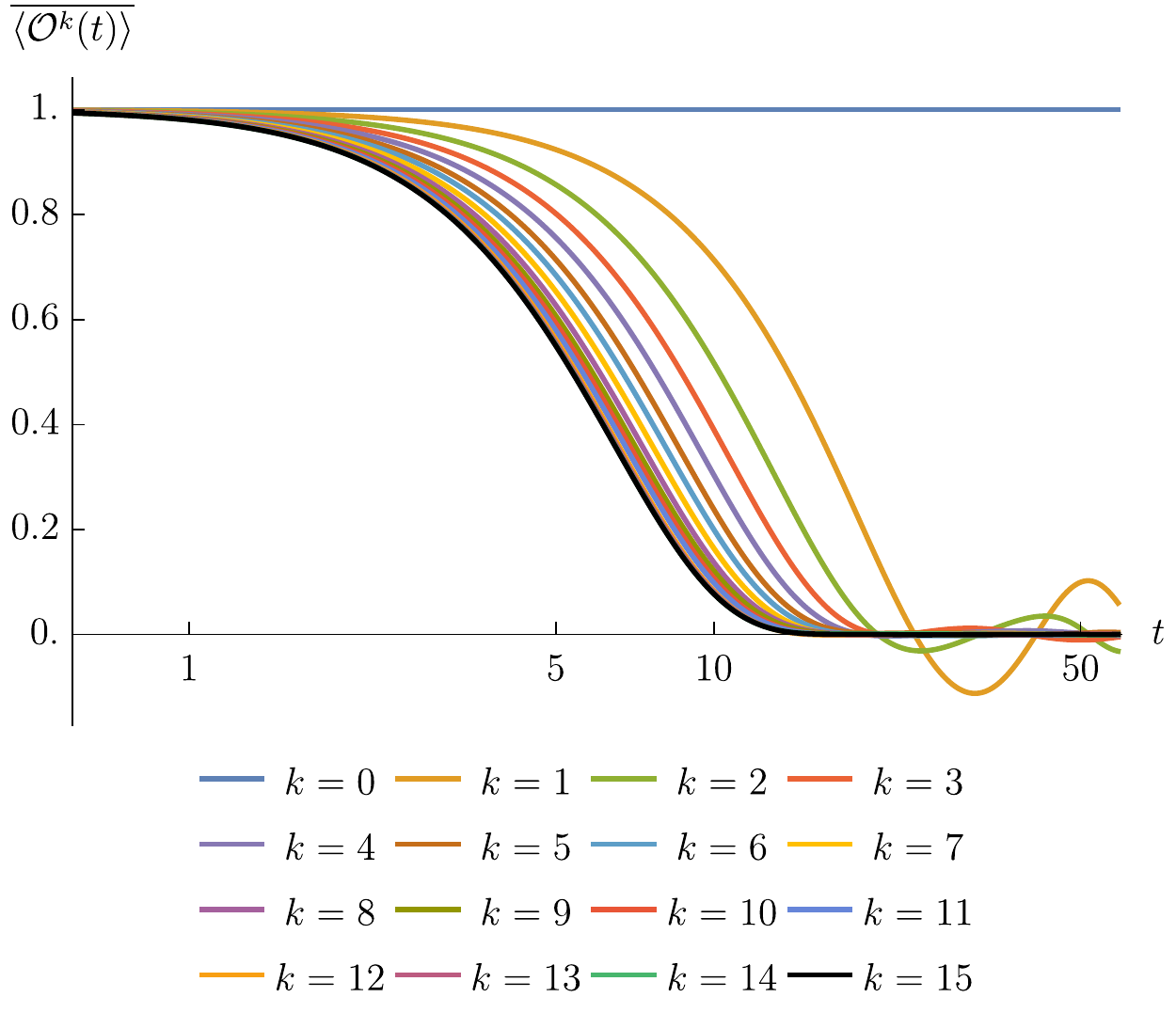}
	\caption{The decay of the correlation functions, $\overline{\langle \mathcal{O}^k (t) \rangle}$, at various values of $k$, for the SYK$_2$ model (with $a = x$) and for the case $L = 15$. The ensemble averages are performed over $300$ ensemble realizations.}
	\label{fig:correlation_decay_SYK2}
\end{figure}

To numerically confirm the above argument, we have studied the time evolution of various spin operators $\overline{\langle\mathcal{O}^k(t)\rangle}$ in the $a=x$ model. We have checked (not shown) that the evolution of the return amplitude turns out to be solely characterized by the operator size $k$. For instance, the decay patterns of $\overline{\langle \hat \sigma_i^x \hat \sigma_{i+\ell}^x(t) \rangle}$ are essentially identical across all $1 \leq i \leq L - \ell$, while they strongly depend on the value of $\ell$. Furthermore, Fig.~\ref{fig:correlation_decay_SYK2} exhibits the pattern that the operators with higher $k \leq L$ decay faster to zero, in agreement with our prediction. We also confirmed (not shown) that a greater value of $k>L$ leads to a slower decay rate, since the dimension \eqref{eq:dim} of the operator space shrinks as the size $k>L$ gets bigger.

The relation between the size of the operator and the decay rate suffices to explain how the connected part of the spin-spin correlation functions \eqref{eq:magnetic_suscept_ij}
evolve in time. In the $a = z$ model, the first term, which is a correlator of size $4$, vanishes slightly slower than the second term, which is the {\it product} of two size~$2$ operators. The difference between the two terms is small and independent of the choice of $i$ and $j$, in agreement with Fig.~\ref{fig:histogram}.

On the contrary, in the $a = x$ model,  $\chi_{i j}^{x }(t)$ is the difference between a correlator of size $2(j - i)$ and the product of two correlators of respective sizes $(2 j - 1)$ and $(2 i - 1)$.
The peak height $P_{ij}^x$ is therefore maximized when $i$ and $j$ are adjacent, \textit{i.e.} $j = i \pm 1$, \textit{and} $i  \sim j \sim L/2$.
Furthermore, since the product of correlators diminishes much more rapidly than the single correlator term, $P_{ij}^x$ is generically larger than $P_{ij}^z$, thereby contributing to the bigger bump of $\chi^x(t)$, as visualized in Fig~\ref{fig:two_point}.
It is remarkable that, although the initial Hamiltonians, \eqref{eq:H_0_def} and \eqref{eq:SYK_hamiltonians_def}, do not explicitly involve any couplings between neighboring spins,  the notion of locality on the spin lattice emerges from the Majorana representation of the spin operators, due to the operator hopping dynamics under the SYK${}_2$ quench and by averaging over the Gaussian couplings.

An important point to stress is that the fast decay of the averaged correlator is a consequence of the property that the interactions are all-to-all, which remains true for any SYK$_q$ models irrespectively of the value of $q$.
In the current case of the SYK${}_2$ model, thanks to all-to-all interactions, the operator  $\mathcal{O}^k(t)$ can sweep across the full space of size $k$ operators, leading to the fast decay of the return amplitudes.
More generally, the condition for return amplitudes to decay is that the space that $\mathcal{O}^k$ can explore under the unitary time evolution is large enough, which can be satisfied whenever the operator $\mathcal{O}^k$ sits on a connected (hyper)-graph \cite{Chen:2019klo, Lucas:2019cxr}.

\section{From operator hopping to operator growth}
\label{sec:operator_growth}

In this section, we inspect how the evolution of $\overline{\langle \mathcal{O}^k (t) \rangle}$ after a quantum quench protocol can discriminate between two different types of operator dynamics: operator hopping, discussed in section \ref{sec:k-locality}, versus {\it operator growth}, discussed, for example, in \cite{Roberts:2018mnp}. To this end, we turn now to the case in which the quench Hamiltonian is the SYK$_q$ Hamiltonian with $q > 2$ and we focus on the case $q = 4$.

Contrary to the $q = 2$ algebra, \eqref{eq:operator_hopping_algebra}, which preserves the operator size, the commutation relation at $q=4$,
\begin{equation}
\label{eq:operator_growth_algebra}
\left[\mathcal{\hat H}_{1}^{(4)} , \hat \gamma^i \right] = \sum_{j < k < l} J_{ijkl} \, \hat \gamma^j   \hat \gamma^k  \hat \gamma^l \ ,
\end{equation}
realizes an example of the operator growth dynamics; the evolution of a fundamental operator $\hat\gamma^i$ is not confined in the space of size 1 operators, but its
degree of locality continues to increase with time until it saturates $L$.
Notice that the notion of the operator growth is a hallmark of the early-time quantum chaos.
The underlying idea is that the operator growth dynamics can develop simple (even fundamental) operators into more complex and extended ones, thereby realizing a quantum analogue of the well-known ``butterfly effect'' \cite{Sekino_2008, Shenker:2013pqa}.

We recall that the decay rate of $\overline{\langle \mathcal{O}^k (t) \rangle}$ is governed by the dimension of the operator space reached out from an initial $k$-local observable $\mathcal{O}^k$ through the time evolution.
This implies that the decay rate of $\overline{\langle \mathcal{O}^k (t) \rangle}$ should be less sensitive to the value of $k$ under the operator growth, compared to the hopping dynamics.
As explained before, the operator hopping {\it does not} change the operator size, therefore a set of possible trajectories is also confined in the space with the fixed dimension \eqref{eq:dim}. Under operator growth, however, the degree of locality of time-evolved operator quickly saturates to $L$ regardless the initial value of $k$. Since the dimension \eqref{eq:dim} of the operator space with a fixed size $k$ is maximized at $k=L$, the Gaussian averaged correlator $\overline{\langle \mathcal{O}^k (t) \rangle}$ should vanish in a shorter time span, as well as being less sensitive to $k$.

\begin{figure}[t]
\centering
	\includegraphics[width=8cm]{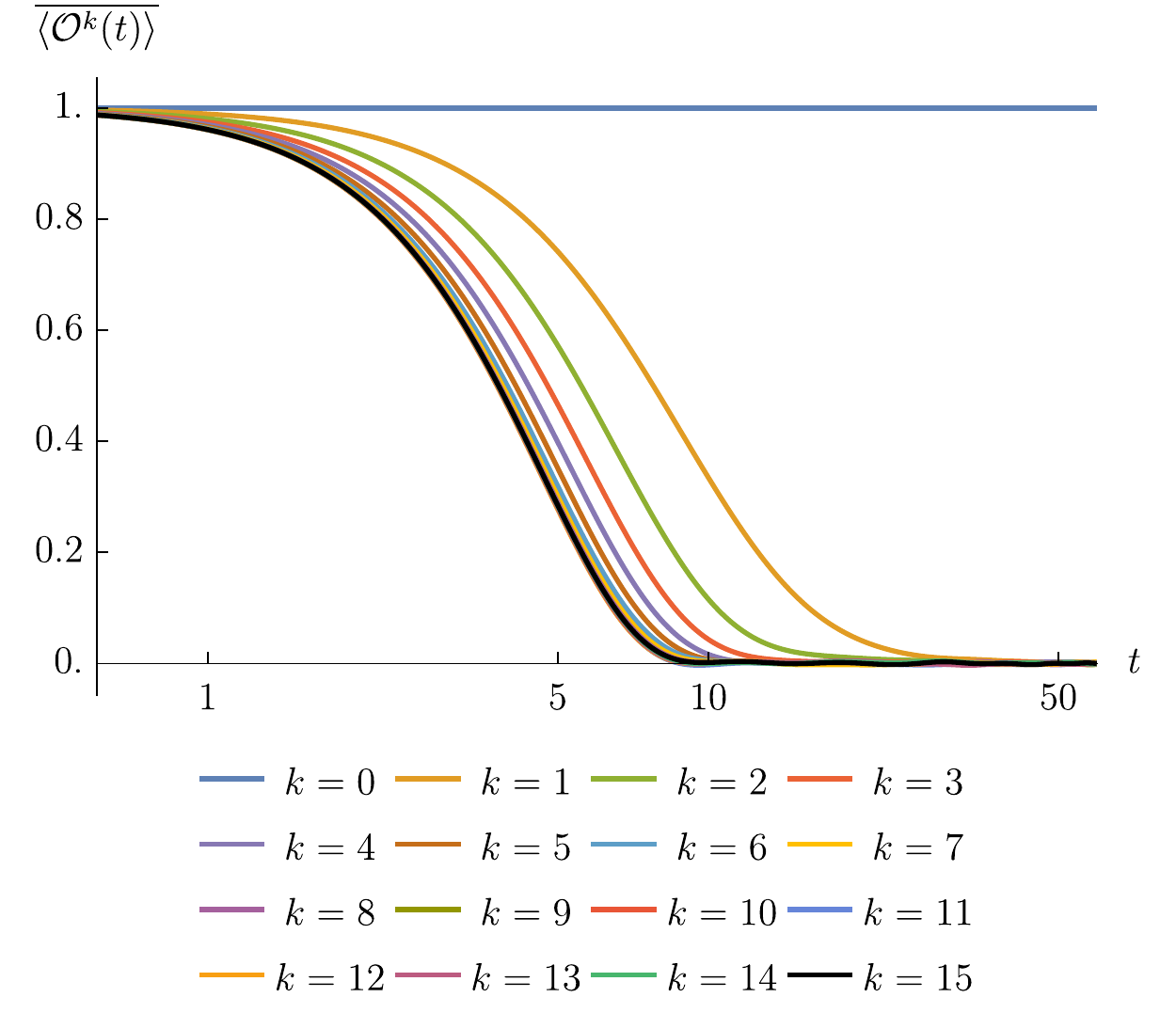}
	\caption{The decay of the correlation functions, $\overline{\langle \mathcal{O}^k (t) \rangle}$, at various values of $k$, for the SYK$_2$ model (with $a = x$) and for the case $L = 15$. The ensemble averages are taken over $100$ realizations.}
	\label{fig:correlation_decay_SYK4}
\end{figure}
To validate the above reasoning, we have computed the averaged correlators under the SYK$_4$ quench and plotted them into Fig.~\ref{fig:correlation_decay_SYK4}.
By contrasting Fig.~\ref{fig:correlation_decay_SYK4} with Fig.~\ref{fig:correlation_decay_SYK2}, which displays the averaged correlators under the SYK$_2$ quench, we recognize that the decay curves start to overlap, becoming indistinguishable from each other, at a smaller values of $k$ in the case of SYK${}_4$ quench. This is consistent with the argument just presented.

\begin{figure}[tbp]
\centering
	\includegraphics[width=8cm]{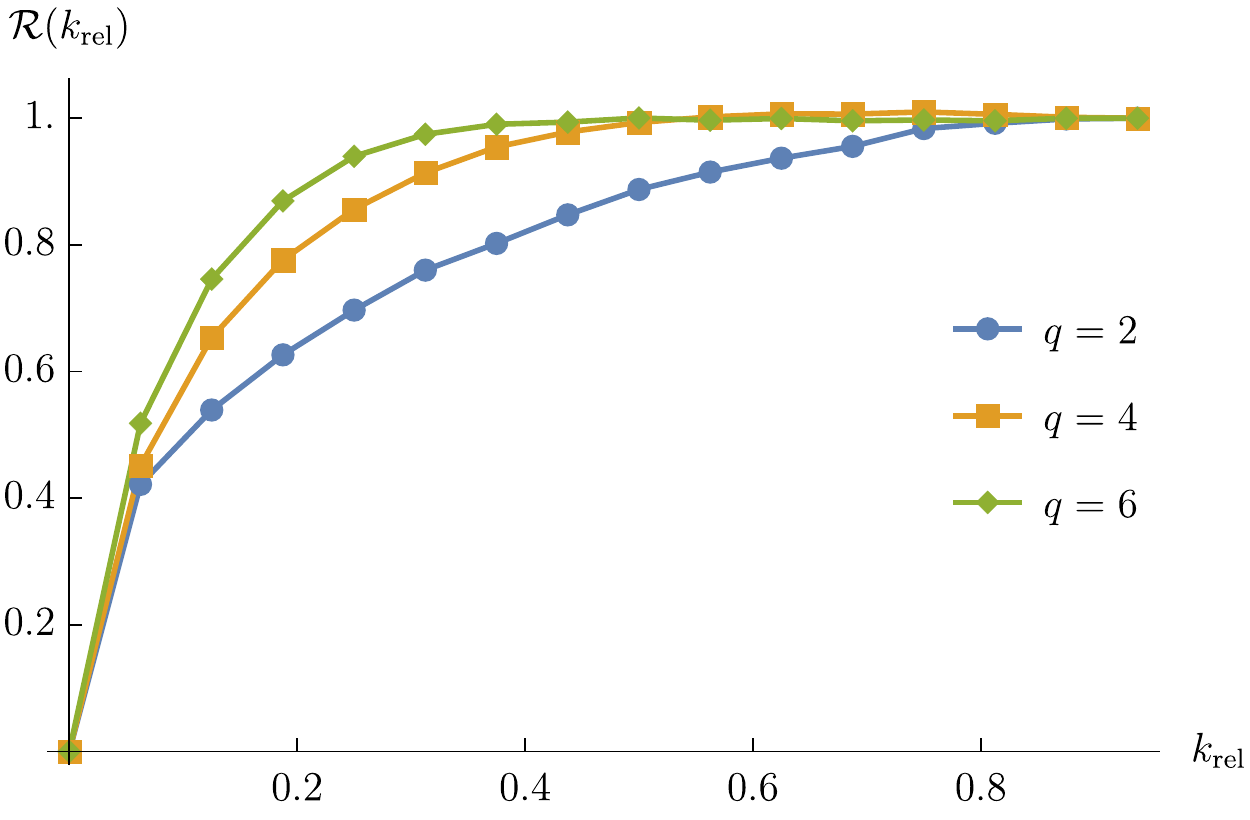}
	\caption{$ \mathcal{R}(k_\mathrm{rel})$, as function of $k_\mathrm{rel}$ for $L = 15$ and for $q = 2 , \, 4$ and $6$.}
	\label{fig:max_derivative_decay_N30}
\end{figure}

To better distinguish the decay pattern of the  correlators, $\overline{\langle \mathcal{O}^k (t) \rangle}$, under the SYK$_4$ vs SYK$_2$ quench dynamics, one can directly look at the decay rate,
\begin{equation}
    \mathcal{D}_{k} (t) \equiv \left| \frac{\mathrm d}{\mathrm{dt}} \overline{\langle \mathcal{O}^{k} (t) \rangle} \right| ,
\end{equation}
focusing on its maximum value, $\max_t (\mathcal{D}_{k} (t))$, the highest speed of correlation decay over time. As we compare the maximum decay rate across different SYK${}_q$ models and various sizes $L$ of the spin lattice, it is also convenient to normalize the degree $k$ of locality as $ k_\mathrm{rel} \equiv k / L$,
and the maximum decay rate, $\max_t (\mathcal{D}_{k} (t))$, as
\begin{equation}
    \label{eq:dmax}
   \mathcal{R}(k_{\text{rel}}) \equiv \frac{\max_t (\mathcal{D}_{k} (t))}{ \max_t (\mathcal{D}_{L} (t))} \ .
\end{equation}
In Fig. \ref{fig:max_derivative_decay_N30}, we display the normalized maximum value \eqref{eq:dmax} of the decay rate as a function of $k_\mathrm{rel}$ for different SYK${}_q$ models. We have found that these plots are insensitive to the lattice size $L$, only depending on the value of $q$.

We observe that $\mathcal{R}(k_{\text{rel}})$ saturates at a smaller value of $k_\text{rel}$ under the SYK$_4$ quench than in the SYK$_2$ case.
This is in agreement with the expected difference between operator growth and hopping: since the hopping dynamics preserves the size of operators, the associated decay rate must be more sensitive as explained above.

For comparison, we also depict the values of $\mathcal{R}(k_{\text{rel}})$ for the SYK$_6$ model,
for which we observe that a saturating $k_{\text{rel}}$ is even smaller than the one for the SYK$_4$ dynamics.
The smaller saturating $k_{\text{rel}}$ for the SYK$_6$ model is easy to explain, since the sextic SYK Hamiltonian is faster in increasing the operator size than the quartic SYK Hamiltonian.
Also, it is worth to notice that the difference between the SYK$_4$ and SYK$_6$ models is much less pronounced than that between the SYK$_4$ and SYK$_2$ models.

The above numerical results strongly suggest that the maximum value $\mathcal{R}(k_{\text{rel}})$ of the decay rate of the averaged spin correlators can effectively distinguish the dynamics of operator growth versus hopping. However, from its definition, $\mathcal{R}(k_{\text{rel}})$ is not a convenient quantity for direct experimental access.
Indeed, to distinguish between hopping and growth, one still should in principle  measure the correlators $\overline{\langle \mathcal{O}^{k} (t) \rangle}$ for all the possible values of $k$, from $k = 1$ to $k = L$.
Considering large spin-chains, this would imply the necessity of measuring a very large number of \textit{averaged} correlators, thus making the requirements in terms of number of measurements very demanding.

Given the considerations, we ask whether there is a simple correlation function, which alone can discriminate between operator hopping and growth.
As it can be inferred from Fig.~\ref{fig:max_derivative_decay_N30}, the averaged correlators at large degree of locality are not very useful
since both hopping and growth are in the saturation regime at large $k$ and the resulting dynamics will be indistinguishable.
Instead, the difference between two kinds of dynamics must be evident at small degree of locality, such as $k = 1$ or $k=2$. It is therefore  interesting to study the correlator $\chi_{1L}^x$, involving the difference between an operator of size $2(L-1)$, whose dynamics is identical to the operator of initial size $2$, and the product of two operators of effective size $1$.
 Hence, we expect to observe the maximum difference between hopping and growth using this probe.
\begin{figure}[t]
\centering
	\includegraphics[width=8cm]{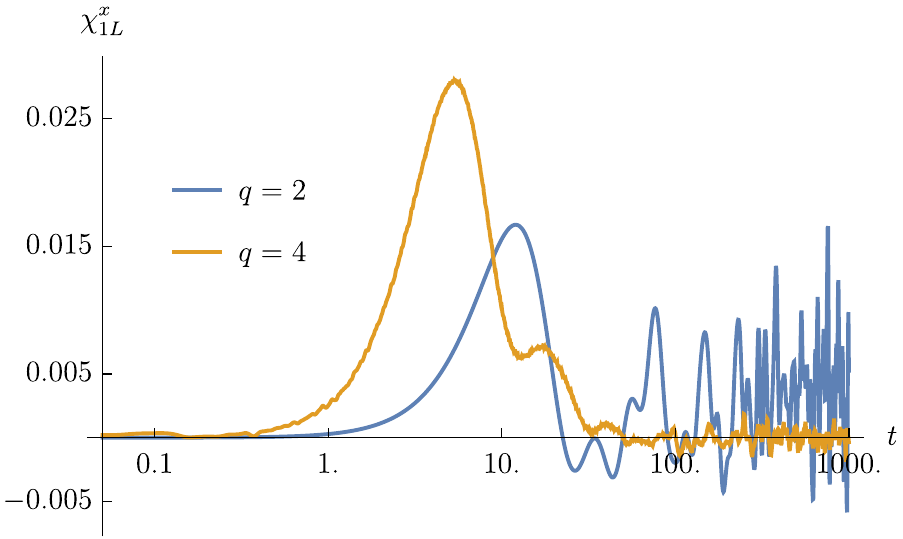}
	\caption{The spin-spin correlators, $\chi^x_{1L}$, computed at $L = 30$ and averaged over $100$ ensemble realizations, for both the $q = 2$ and the $q = 4$ cases.}
	\label{fig:chi_1L_q2_q4}
\end{figure}

Fig.~\ref{fig:chi_1L_q2_q4} contrasts the time evolution of the correlator $\chi^x_{1L}(t)$ in the $q = 2$ and the $q = 4$ models.
The ensemble averages are taken over $100$ sets of random couplings. It is immediate to notice that the behaviors are rather different. In SYK${}_4$ model, the correlation function exhibits a clear maximum, higher than the noise amplitude at late times by a few order of magnitudes; \textit{i.e.} the peak is by far larger than the late time fluctuations, still present after the average over the Gaussian couplings, although highly suppressed.
On the other hand, in SYK${}_2$ model, the height of the initial peak is of the same order of the fluctuation amplitude at late times; the SYK${}_2$ fluctuations are more pronounced than the SYK${}4$ noises, since in the case of operator hopping the evolution trajectory of the given operator is confined to the relatively smaller space of operators of size $2(L-1)$.

We have extensively checked that these marked differences are robust by increasing the number of ensemble realizations over which the Gaussian average is performed.
 Moreover, they become parametrically more evident by increasing lattice size $L$. These observations suggest that the evolution of $\chi^x_{1L}(t)$ in time can be considered as a useful diagnostics of operator growth versus hopping, being able to identify the nature of the operator size dynamics.

\section{Conclusions}
\label{sec:conclusions}

In this paper, we have investigated whether quantum chaos, and specifically operator growth, can be revealed by performing quantum quench protocols on systems defined over spin lattices.

By using the celebrated SYK$_q$ model as the quench Hamiltonian with all-to-all interactions,
we have established that the time evolution of the spin-spin correlation functions can be used as a probe of operator growth.

Mapping the spin variables to the Majorana fields, which here constitute the fundamental operators,  the associated size of different spin-spin correlation functions have been identified.

We have demonstrated how the decay rate of the averaged spin correlators $\overline{\langle \mathcal{O}^k (t) \rangle}$ is controlled by their initial sizes.
Moreover, the relative decay rate, $\mathcal{R}(k_{\text{rel}})$, can distinguish operator growth from operator hopping; the former being a hallmark of early-time quantum chaos while the latter shows a rather trivial dynamics in operator space.
Finally, we have discussed that the difference in operator dynamics strongly affects the particular averaged spin correlator $\chi^x_{1L}(t)$: the amplitude of the late-time fluctuations is comparable to the height of an initial peak under operator hopping, but significantly suppressed under operator growth.
Such marked distinction can be useful to detect quantum chaos in strongly correlated systems, especially given the fact that the evolution of the spin correlators is experimentally traceable by using state-of-the-art imaging techniques and quantum gas microscopy \cite{Kan_sz_Nagy_2018}.
We believe that a precise analysis of the connections between the peaked structure that emerges in the dynamics of spin-spin correlation functions  $\chi^x_{1L}$ and the OTOCs is worth of future investigations.
In particular it would be extremely interesting to understand whether a quantitative evaluation of the Lyapunov exponents can be extracted from $\chi^x_{1L}$.
Also, it would be intriguing to analyze the evolution of averaged spin correlators in other chaotic systems, \textit{e.g.} random circuits \cite{Bentsen6689, Hartmann:2019cxq}, from the perspective of operator dynamics.

\section*{Acknowledgement}
We thank F.~Haehl, T.~Nosaka, V.~Rosenhaus for related discussions.
JK acknowledges the support from the NSF grant PHY-1911298. DR is supported by a KIAS Individual Grant PG059602 at Korea Institute for Advanced Study. Most numerical computations were done thanks to the computing resources provided by the KIAS Center for Advanced Computation (Abacus System) and the Institute for Advanced Study.
Some of the numerical results have been obtained by making use of the Wolfram Mathematica package \verb|QuantumManyBody|, freely available on \href{https://github.com/Dario-Rosa85/QuantumManyBody}{GitHub}. The rendering of the plots has been realized using the the Wolfram Mathematica package \verb|MaTeX|, freely available on the \href{https://library.wolfram.com/infocenter/MathSource/9355/}{Wolfram Library Archive}.

\appendix

\section{Bosonic models}
\label{App:hard-core_bosons}

In this appendix, we investigate the real hard-core boson variants of the SYK models. We study how their dynamics differ from their fermionic counterparts. We recall that the \textit{complex} hard-core boson SYK model was introduced and studied in \cite{Fu_2016, Iyoda_2018}.
Despite the same symmetry breaking pattern as in the fermionic models, the bosonic dynamics turns out to be completely different.
This difference can be understood again in terms of the size of the operators involved.

To begin with, we define the real hard-core boson operators, $\hat \chi^i$, with $i  = 1, \dots , 2L $, as the operators satisfying the following algebra:
\begin{align}
\label{eq:bosons_algebra}
\{ \hat \chi^{a} , \, \hat \chi^{b} \} &=
\begin{cases}
    0 & \text{if }(a, b)=(2i-1, 2i) \text{ for } 1 \leq i\leq L\\
    1 & \text{if }a=b
\end{cases} \nonumber\\
\big[ \hat \chi^a , \, \hat \chi^b \big]  &=
\begin{cases}0 & \mathrm{otherwise}. \end{cases}
\end{align}
The hard-core bosonic operators can be mapped to operators defined on a spin lattice of length $L$ via the following JW-like map
\begin{align}
\label{eq:JWmap_bosons}
& \hat \chi^{2j-1}  =  \frac{1}{\sqrt{2}} \, \hat \sigma^x_j \ , \nonumber \\
& \hat \chi^{2j}   =  \frac{1}{\sqrt{2}} \, \hat \sigma^y_j \ .
\end{align}

Similarly to the fermionic case, one can define the following SYK-like Hamiltonian for the operators $\{ \hat \chi^i \}$,
\begin{equation}
\label{eq:SYK_hamiltonians_def_bosonic}
\mathcal{\hat H}_{1, \, \mathrm{B}}^q =  \sum_{i_1 < \dots < i_q} (i)^{s(i_1 , \dots , i_q)} \,  J_{i_1 \dots i_q} \hat \chi^{i_1} \dots  \hat\chi^{i_q} \ ,
\end{equation}
where the coupling constants $J_{i < j < k < l}$ are sampled from the same Gaussian distribution as in \eqref{eq:Jcouplings_variances}.
The factor $s(i_1 , \dots , i_q)$ reads
\begin{equation}
    \label{eq:sign_function_bosons}
s(i_1 , \dots , i_q) \equiv \sum_{ l = 2 }^{q}  (1 + (-1)^{i_l}) \, \delta_{i_{l - 1} + 1 ,\,  i_l} \ ,
\end{equation}
and it must be introduced to ensure that $ \mathcal{\hat H}_{1, \, \mathrm{B}}^q$ is Hermitian under the bosonic algebra \eqref{eq:bosons_algebra}.

Here, for sake of brevity, we focus on the $q = 4$ model. We contrast the temporal evolution of averaged correlators under the quench protocol \eqref{eq:quench} with the Hamiltonian either being the bosonic SYK${}_4$ \eqref{eq:SYK_hamiltonians_def_bosonic} or its fermionic counterpart \eqref{eq:SYK_hamiltonians_def}.
Notice that the symmetry breaking pattern in the bosonic models is the same as in the fermionic systems; in the $a = x$ model the spin symmetry along the $x$ axes is fully broken by the quench term, while in the $a = z$ model the spin symmetry along the $z$ axes is partially broken to the chiral symmetry.

Let us start by considering the connected part of the two-point function, defined in \eqref{eq:magnetic_suscept}.
As we have shown in the main text, the time evolution of $\chi^a(t)$ as well as individual summands $\chi_{ij}^a (t)$ is controlled by the initial size of the operators in $ \sigma_i^a \sigma_j^a$ and $\sigma_i^a$.
A rapid inspection, by making use of the JW-like map \eqref{eq:JWmap_bosons}, shows that in the $a = x$ case, the bosonic $\sigma_i^x \sigma_j^x$ operators have size $2$, while the bosonic $\sigma_i^x$ operators have size $1$. For the $a = z$ case, however, the operator size is exactly the same as in the fermionic case: the bosonic $\sigma_i^z \sigma_j^z$ operators have size $4$, while the bosonic $\sigma_i^z$ operators have size $2$.
\begin{align}
    \begin{tabular}{c|cccc}
         Operator
         & $\hat \sigma_i^{z}$
         & $\hat \sigma_i^{z} \hat \sigma_j^{z}$
         & $\hat \sigma_i^{x}$
         & $\hat \sigma_i^{x} \hat \sigma_j^{x}$ \\\hline
         Size
         & $2$
         & $4$
         & $1$
         & $2$
    \end{tabular}
    \end{align}
It is important to note that, in the bosonic case, all the operator sizes are independent of $i$ and $j$.

\begin{figure}
    \centering
    \includegraphics[width=8cm
]{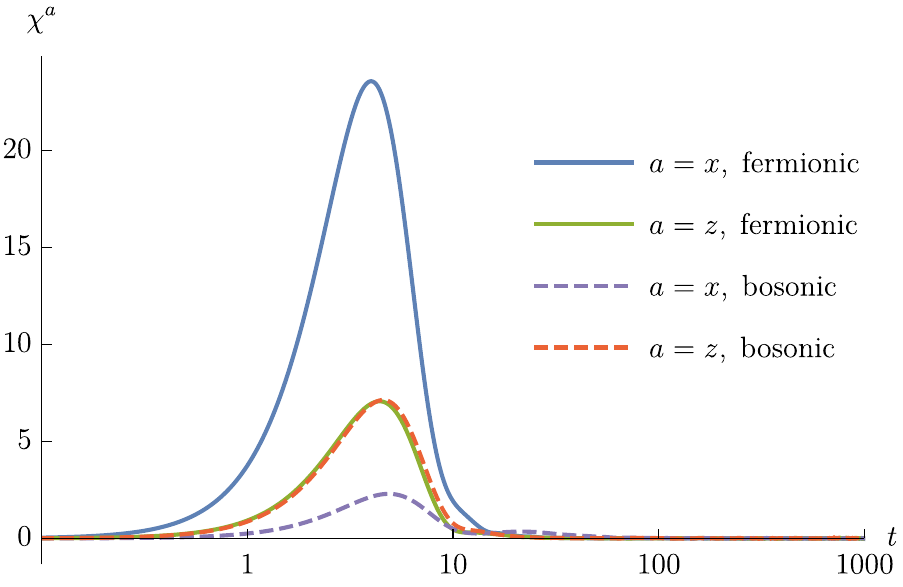}
    \caption{The function $\chi^a (t)$, with $a = x, \, z$ for both the bosonic and fermionic models computed for $L = 15$ and with $q = 4$. The ensemble averages are taken over $300$ realizations.}
    \label{fig:quartic_susceptibility}
\end{figure}

\begin{figure}[]
    \centering
    \includegraphics[width=\linewidth]{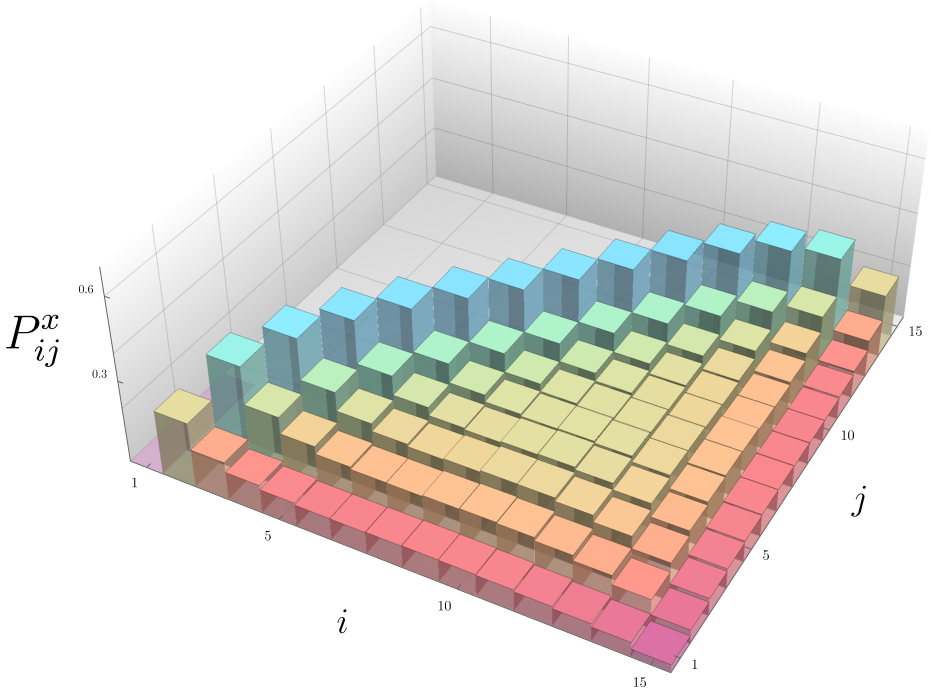}
    \includegraphics[width=\linewidth]{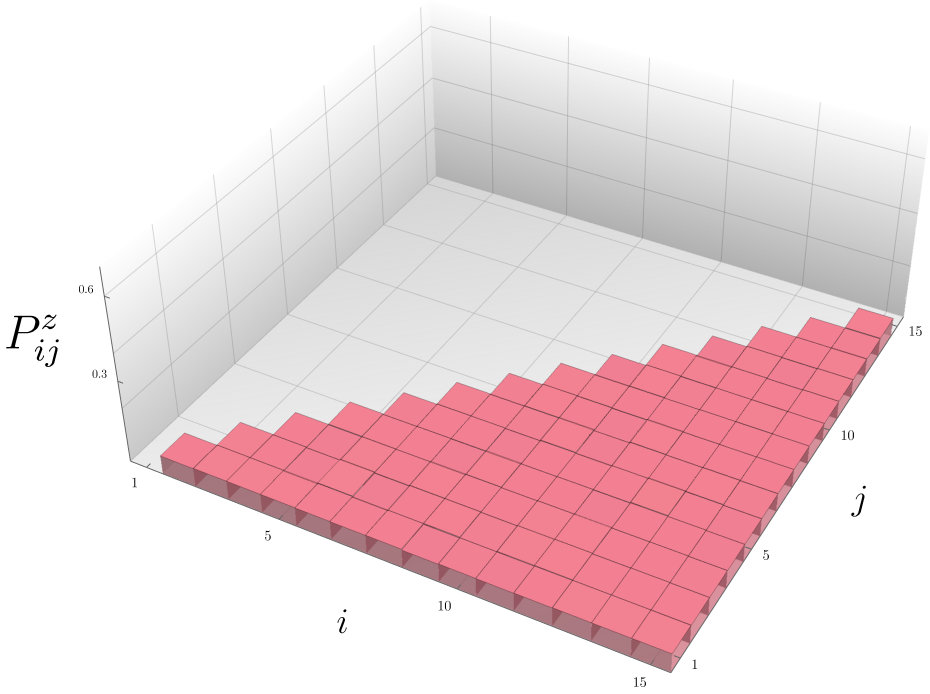}   \caption{The peak heights of $P^x_{ij} $ and $P^z_{ij} $ in fermionic SYK${}_4$ models, as a function of the lattice sites $i,j$ at $L=15$ and averaged over $100$ ensemble realizations.}
    \label{fig:hist-syk4-f}
\end{figure}

\begin{figure}[]
    \includegraphics[width=\linewidth]{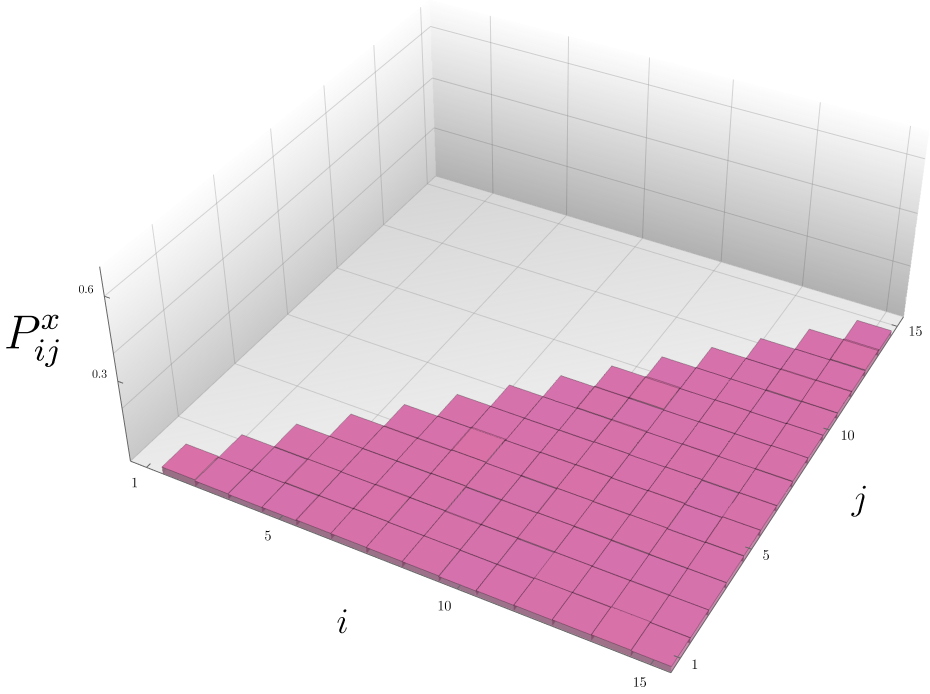}
    \includegraphics[width=\linewidth]{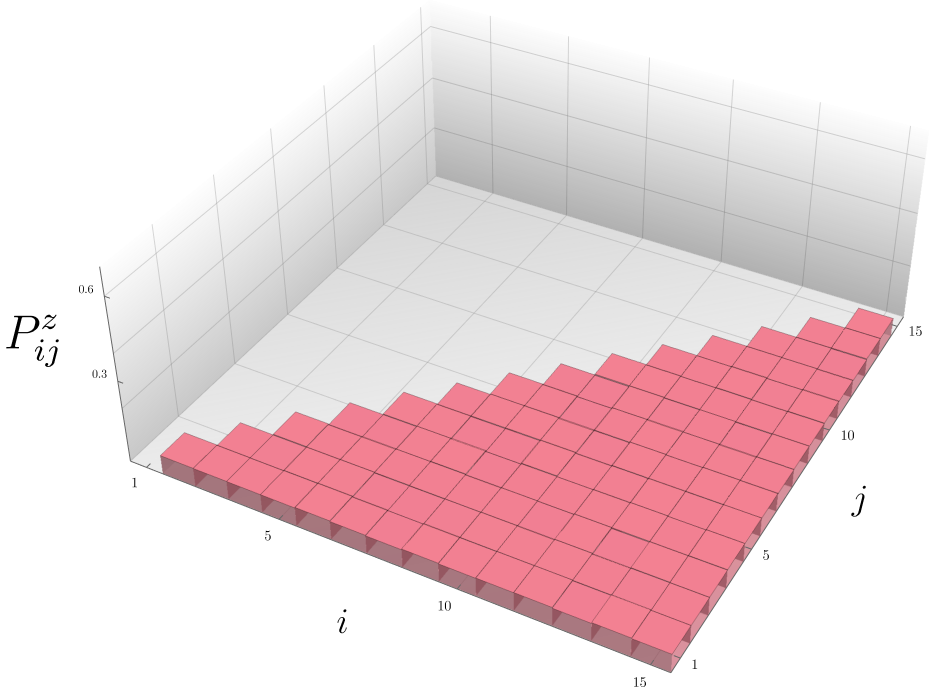}
    \caption{The peak heights of $P^x_{ij} $ and $P^z_{ij} $ in bosonic SYK${}_4$ models, as a function of the lattice sites $i,j$ at $L=15$ and averaged over $100$ ensemble realizations.}
    \label{fig:hist-syk4-b}
\end{figure}

The correlators $\chi^a(t)$, for bosonic/fermionic SYK${}_4$ models are compared in Fig.~\ref{fig:quartic_susceptibility}.
We clearly observe that, for $a = z$, the evolution curves for both bosonic and fermionic models are barely distinguishable; this shows a perfect agreement with the expectation based on the size of the operators involved, since when $a = z$ the size is independent of the bosonic/fermionic nature of fundamental variables.
On the other hand, the situation is completely different for $a = x$ case; the characteristics for the fermionic and bosonic models are rather opposite, where the fermionic/bosonic models exhibit a huge/tiny value of the peak, respectively.
Such distinction is again in perfect agreement with the expectation based on the operator size, because the fermionic $a = x$ model is the only case for which we have very large differences in the operator sizes between $\sigma_i^x \sigma_j^x$ and $\sigma_i^x $.
It is also interesting to observe that the fermionic $a = x$ model is the only model for which the height of the peak, $\chi^x_\mathrm{max} (L)$, scales quadratically as a function of $L$ (Fig~\ref{fig:two_point}). We also display  the height $P_{ij}^a$ of the peak for all bosonic/fermionic SYK${}_4$ variants in Figs~\ref{fig:hist-syk4-f}~and~\ref{fig:hist-syk4-b}.

\section{The average over the Gaussian coupling}
\label{App:gaussian_average_perturbative}

Here we present an analytic argument, based on a short time expansion, to establish a relation between the degree of operator locality and the decay rate of correlation functions.
As a product, we show that the numerical results presented in Fig.~\ref{fig:correlation_decay_SYK2} and Fig.~\ref{fig:correlation_decay_SYK4} indicate that the correlators $\overline{\langle \mathcal{O}^k (t) \rangle}$ are computing a return amplitude, as the effect of the average over the Gaussian couplings.

Thanks to the Majorana representation, \eqref{eq:k-locality_z_system} and \eqref{eq:k-locality_x_system}, of the spin variables, it is easy to consider the time-evolution of spin operators.
 One needs to resolve the commutator $[\mathscr{\hat H}_1, \mathcal{O}^k(t)]$ that appears in the Heisenberg equation of motion \eqref{eq:Heisenberg}  by repeated applications of the following commutation relation:
\begin{align}
    \label{eq:comm-rel}
    [\mathcal{\hat H}_{1}^{(q)} , \hat \gamma^n ] = \mathrm{i}^{q/2}\sum_{\substack{i_1 < \cdots <i_{q-1}\\i_1, \cdots, i_{q-1} \neq n}}  J_{i_1 \cdots i_{q-1} n} \, {\hat \gamma}^{i_1}\cdots {\hat \gamma}^{i_{q-1}}
\end{align}
between the SYK${}_q$ coupling \eqref{eq:SYK_hamiltonians_def} and a Majorana operator.
For notational convenience, we assume the quench Hamiltonian $\mathscr{\hat H}_1$ to be the standard SYK$_q$ Hamiltonian $\mathcal{\hat H}_{1}^{(q)}$ without normalizing the bandwidth. Instead, the bandwidth normalization has been directly taken into account as a proper rescaling of the time variable $t$ in \eqref{eq:time-evolution-klocal}.

Since the post-quench Hamiltonian is static, the solution $\mathcal{O}^k(t)$ of the Heisenberg equation can be written as:
\begin{align}
    \label{eq:time-evolution-klocal}
    \mathcal{O}^k(t) &= \mathcal{O}^k + \frac{\mathrm{i}t}{\Delta_{\mathcal{\hat H}^{(q)}_1}} [\mathcal{\hat H}_1^{(q)}, \mathcal{O}^k ]+{\frac {(\mathrm{i}t)^2}{2! \Delta_{\mathcal{\hat H}^{(q)}_1}^2}}[\mathcal{\hat H}_1^{(q)},[\mathcal{\hat H}_1^{(q)}, \mathcal{O}^k ]]\nonumber \\
    &+{\frac {(\mathrm{i}t)^3}{3! \Delta_{\mathcal{\hat H}^{(q)}_1}^3}}[\mathcal{\hat H}_1^{(q)},[\mathcal{\hat H}_1^{(q)},[\mathcal{\hat H}_1^{(q)}, \mathcal{O}^k ]]]\\&
    +{\frac {(\mathrm{i}t)^4}{4! \Delta_{\mathcal{\hat H}^{(q)}_1}^4}}[\mathcal{\hat H}_1^{(q)},[\mathcal{\hat H}_1^{(q)},[\mathcal{\hat H}_1^{(q)},[\mathcal{\hat H}_1^{(q)}, \mathcal{O}^k ]]]] + \mathcal{O}(t^5) \nonumber
\end{align}
where $\mathcal{O}^k \equiv \mathcal{O}^k(0)$. Let us replace all the iterated commutators in \eqref{eq:time-evolution-klocal} by using the commutation relation \eqref{eq:comm-rel}.
Averaging over all Gaussian coupling constants, $J_{i_1 \cdots i_{q}}$, leads to huge simplification. All odd-order moments of Gaussian couplings drop out, implying $\overline{\mathcal{O}^k(t)}$ is an even polynomial in $t$. The remaining even-order moments can be handled with Wick's theorem, enforcing the sum of surviving operators on the right-hand side to be proportional to $\mathcal{O}^k(0)$. This shows that the averaged correlator $\langle \overline{\mathcal{O}^k(t)} \rangle$ is indeed a \textit{return amplitude}, \textit{i.e.} it computes the amplitude that  $\mathcal{O}^k(t)$  returns to itself, $\mathcal{O}^k(0) $. Also, it manifests that the early-time scale, which determines the regime of validity of the expansion \eqref{eq:time-evolution-klocal}, is inversely proportional to the standard deviation $J$ of the SYK coupling constants.

\begin{figure}[ht]
\centering
	\includegraphics[width=8cm]{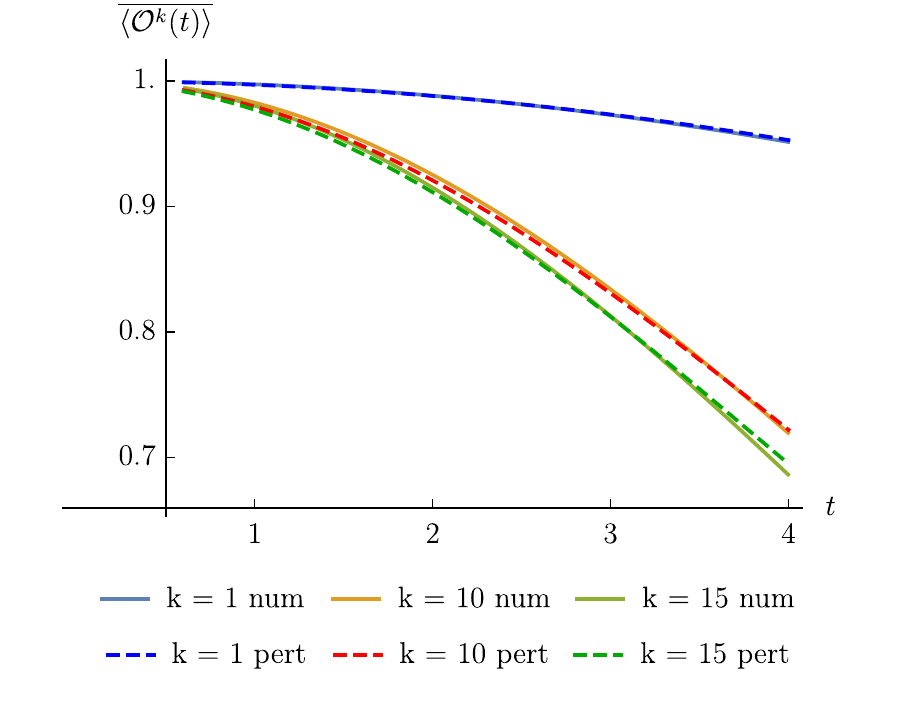}
	\caption{The short time decay of the correlation functions, $\overline{\langle \mathcal{O}^k (t) \rangle}$, at various values of $k$, for the SYK$_2$ model (with $a = x$) and for the case $L = 15$, against its analytic quartic prediction, based on \eqref{eq:2nd-order} and \eqref{eq:4th-order}.}
	\label{fig:correlation_decay_SYK2_analytic}
\end{figure}

We specialize to a simple case with $q=2$ and $k=1$ as an illustrative example. The commutator action $[\mathcal{\hat H}_1^{(2)}, \, \cdot\,  ]$  defined with SYK$_2$ Hamiltonian preserves the size of an operator. Considering only the terms that involve an even number of the commutator action, we find
\begin{align}
\label{eq:2nd-order}
&[\mathcal{\hat H}_1^{(2)},[\mathcal{\hat H}_1^{(2)}, {\hat \gamma}^n ]] = - \sum_{\substack{a, b, c,d \\ a \neq b, c \neq d}} J_{ab}J_{cd}\delta_{bn}\delta_{da} \cdot \hat{\gamma}^c,\\[0.2cm]
&[\mathcal{\hat H}_1^{(2)},[\mathcal{\hat H}_1^{(2)},[\mathcal{\hat H}_1^{(2)},[\mathcal{\hat H}_1^{(2)}, {\hat \gamma}^n ]]]] = \nonumber \\[0.05cm]& \sum_{\substack{a, b, c,d,e,f,g,h \\ a \neq b, c \neq d,e \neq f, g \neq h}} J_{ab}J_{cd}J_{ef}J_{gh}\delta_{bn}\delta_{da}\delta_{fc}\delta_{he} \cdot \hat{\gamma}^g,
\label{eq:4th-order}
\end{align}
and so on.
The right-hand side expressions are further simplified by taking the coupling constant average,:
\begin{align}
    \label{eq:2nd-moment}
    \overline{J_{ab}J_{cd}} &= \frac{J^2}{2L} (\delta_{ac}\delta_{bd} - \delta_{ad}\delta_{bc}),\\
    \overline{J_{ab}J_{cd}J_{ef}J_{gh}} &= \overline{J_{ab}J_{cd}} \cdot \overline{J_{ef}J_{gh}} \nonumber\\&
    + \overline{J_{ab}J_{ef}} \cdot \overline{J_{cd}J_{gh}}
    \label{eq:4th-moment}
    \\&+\overline{J_{ab}J_{gh}} \cdot \overline{J_{cd}J_{ef}}.\nonumber
\end{align}
Just like \eqref{eq:4th-moment}, the higher-order moments can be obtained by Wick contractions. Inserting them back, the summations appearing in iterative commutators, e.g., \eqref{eq:2nd-order}--\eqref{eq:4th-order}, can be easily carried out.
\begin{align}
\overline{[\mathcal{\hat H}_1^{(2)},[\mathcal{\hat H}_1^{(2)}, {\hat \gamma}^n ]]} &=\frac{(2L-1)J^2}{2L} \hat{\gamma}^n,\\
\overline{[\mathcal{\hat H}_1^{(2)},[\mathcal{\hat H}_1^{(2)},[\mathcal{\hat H}_1^{(2)},[\mathcal{\hat H}_1^{(2)}, {\hat \gamma}^n ]]]]} &=  \frac{(2L-1)(4L-1) J^4}{4L^2}\hat{\gamma}^n \nonumber.
\end{align}
The above expressions can be simply generalized to the followings that hold for spin operators with $k$-locality $k \geq 1$:
\begin{align}
\label{eq:comm-2-syk2}
&\overline{[\mathcal{\hat H}_1^{(2)},[\mathcal{\hat H}_1^{(2)}, \mathcal{O}^k ]]} =\frac{k(2L-k)J^2}{2L} \, \mathcal{O}^k ,\\[0.3cm]
\label{eq:comm-4-syk2}
&\overline{[\mathcal{\hat H}_1^{(2)},[\mathcal{\hat H}_1^{(2)},[\mathcal{\hat H}_1^{(2)},[\mathcal{\hat H}_1^{(2)}, \mathcal{O}^k ]]]]}  \nonumber \\ &\qquad = \frac{k(2L-k) ((6k-2) L-(3k^2-2))J^4}{4L^2}\, \mathcal{O}^k.
\end{align}
and they constitute the solution \eqref{eq:time-evolution-klocal}  of  Heisenberg equation. In Fig.~\ref{fig:correlation_decay_SYK2_analytic}
\begin{figure}[ht]
  \centering
    \includegraphics[width=8cm]{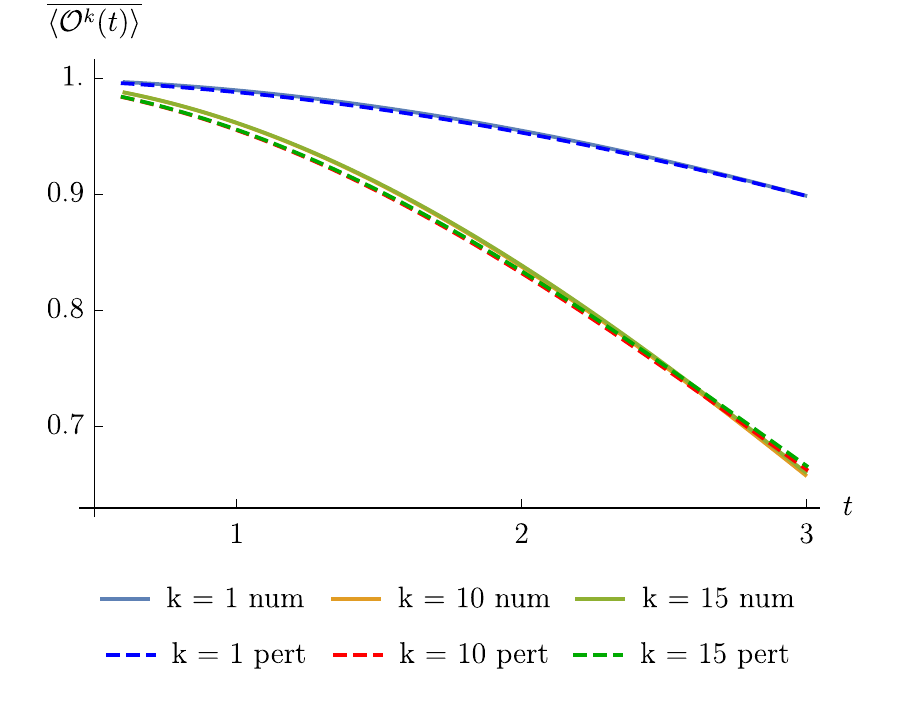}
    \caption{The short time decay of the correlation functions, $\overline{\langle \mathcal{O}^k (t) \rangle}$, at various values of $k$, for the SYK$_4$ model (with $a = x$) and for the case $L = 15$, against its analytic quartic prediction, based on \eqref{eq:comm-2-syk4} and \eqref{eq:comm-4-syk4}.}
    \label{fig:correlation_decay_SYK4_analytic}
  \end{figure}
we compare some representative correlation functions, $\overline{\langle \mathcal{O}^k (t) \rangle}$, against their analytical quartic predictions based on \eqref{eq:2nd-order} and \eqref{eq:4th-order}. From the figure we clearly see that the short time agreement is extremely good, thus confirming that the numerical average over the Gaussian couplings, performed to get the numerical plots of $ \overline{\langle \mathcal{O}^k (t) \rangle} $, is sufficiently precise to convince that the numerical correlators $\overline{\langle \mathcal{O}^k (t) \rangle}$ are effectively computing the return amplitudes of the operators $\mathcal{O}^k$.

The extension to the $q > 2$ case is also straightforward in principle. The main impediment for an actual $q>2$ calculation is the rapid growth of the number of operators that appear in iterative applications of the commutator action \eqref{eq:comm-rel}. Since the right-hand side of the commutator \eqref{eq:comm-rel} contains $C(2L-1, q-1)$ distinct terms, the number of operators appearing in the $n$-times commutator action on $\mathcal{O}^k$ grows very quickly, for $q>2$, as observed in the following $L=5$ example:
\begin{equation*}
    \begin{tabular}{|c|c|c|c|c|c|c|}
     & \multicolumn{3}{c|}{SYK${}_2$} & \multicolumn{3}{c|}{SYK${}_4$} \\\hline
         $n$ & $2$ & $3$ & $4$ & $2$ & $3$ & $4$ \\\hline
         $k=1 $  & $81$ & $657$ & $5121$ & $7518$ & $406980$ & ?\\
         $k=2 $  & $200$ & $2176$ & $21676$ & $8568$ & $482720$ & ?\\
         $k=3 $  & $315$ & $4011$ & $46053$ & $8148$ & $499380$ & ?
    \end{tabular}
\end{equation*}
Even after the Gaussian average has been performed, we find that the commutators \eqref{eq:comm-2-syk2}--\eqref{eq:comm-4-syk2} for  SYK${}_4$ model are
\begin{align}
\label{eq:comm-2-syk4}
&\overline{[\mathcal{\hat H}_1^{(4)},[\mathcal{\hat H}_1^{(4)}, \mathcal{O}^k ]]} =\frac{k (2L-k)J^2}{2^5 L^3} \\&
\quad\times\big((k-1)(k-2)   + (2L-(k+1)(2L-(k+2))\big)  \mathcal{O}^k ,\nonumber
\end{align}
as well as
\begin{align}
\label{eq:comm-4-syk4}
&\overline{[\mathcal{\hat H}_1^{(4)},[\mathcal{\hat H}_1^{(4)},[\mathcal{\hat H}_1^{(4)},[\mathcal{\hat H}_1^{(4)}, \mathcal{O}^k ]]]]} =
    \frac{k  J^4 (2L-k )}{2^{10} L^6} \\&
\quad \times \big(
    2^5(3 k +1) L^5-2^4(15 k ^2+ 38 k) L^4+ 2^3(36 k ^3 \nonumber \\&
\quad +122 k ^2+207 k -99) L^3 -2^2(48 k ^4+168 k^3+543 k ^2\nonumber \\&
\quad +352 k -510) L^2+2^3(9 k ^5 + 21 k ^4 + 168 k ^3 + 88 k ^2 \nonumber\\&
\quad +156 k - 304) L-12 k ^6-336 k ^4-624 k ^2 +864\big) \ , \nonumber
\end{align}
which, compared with the analogous formulas for the SYK$_2$ model, \eqref{eq:comm-2-syk2} and \eqref{eq:comm-4-syk2}, suggest how involved can be the dynamics of the model for $q > 2$.
This is, of course, a manifestation of the fact that for $q > 2$ the dynamics turns from operator hopping to operator growth, and this change is reflected in the cumbersome formulas, \eqref{eq:comm-2-syk4} and \eqref{eq:comm-4-syk4}.
Anyway, we can make use of \eqref{eq:comm-2-syk4} and \eqref{eq:comm-4-syk4} to compare the short time decay of the numerical correlators in the SYK$_4$ model, against their analytical predictions based on \eqref{eq:comm-2-syk4} and \eqref{eq:comm-4-syk4}. The results are reported in Fig.~\ref{fig:correlation_decay_SYK4_analytic}.

Also in this case, an excellent agreement is found at short times.
In particular, it is possible to note that already at the $4$th perturbative order the averaged correlators at large values of $k$ do overlap with each other.


%

\end{document}